# Large Magnetic Anisotropy of an Iron-Porphyrin Complex on Metal Substrate

Bing Liu[1,2], Huixia Fu[1,2], Jiaqi Guan[1,2], Bin Shao[3], Sheng Meng[1,2,4]*, Jiandong Guo[1,2,4]* and Weihua Wang[1]*

[1]Beijing National Laboratory for Condensed Matter Physics and Institute of Physics, Chinese Academy of Sciences, Beijing 100190, China

[2]School of Physical Sciences, University of Chinese Academy of Sciences, Beijing 100190, China

[3]Bremen Center for Computational Materials Science, University of Bremen, Bremen, Germany

[4]Collaborative Innovation Center of Quantum Matter, Beijing 100871, China

*Email: smeng@iphy.ac.cn; jdguo@iphy.ac.cn; weihuawang@iphy.ac.cn

Abstract

Single magnetic atoms or molecules with large single-ion magnetic anisotropy are highly desired for future applications in high-density data storage and quantum computation. Here we have synthesized an Fe-porphyrin complex on Au(111) substrate in a controlled way, and revealed large magnetic anisotropy energy of more than 15 meV by low-temperature scanning tunneling microscopy and spectroscopy. Two magnetic states with opposite spin directions are discriminated by inelastic electron tunneling spectroscopy in varied magnetic field, and are found to have long spin lifetimes. First-principle calculations reveal that the weak ligand filed in this complex keeps the Fe atom in a high-spin state and preserves its large orbital angular momentum, which gives rise to an easy-axis perpendicular to the molecular plane and large magnetic anisotropy energy by spin-orbit coupling.

For the ultimate goal to realize high-density data storage and quantum computation in single atoms or molecules [1,2], there is on-going pursuit of single magnetic atoms and molecules with large magnetic anisotropy [3], which describes the directionality and stability of spontaneous magnetization. When magnetic atoms are adsorbed on metal substrates, their magnetic moments are easily screened or even quenched by itinerant electrons [4,5], and the reported single-ion magnetic anisotropy energy is limited to no more than 10 meV [6-8]. To achieve large magnetic anisotropy, an insulating layer was used to decouple magnetic atoms from metal substrates, whereas this strategy requires the system to be prepared and kept at low temperature [3,9-12]. In organometallic complexes, the molecular ligands provide an alternative way to decouple magnetic atoms from metal substrates [13,14]. Moreover, the ligand field surrounding the magnetic atom can determine its spin and orbit degrees of freedom, and thus dictates its magnetic anisotropy by spin-orbit coupling [15,16]. In this scenario, a weak ligand field that can preserve the atom's orbital angular momentum is highly desired to achieve large magnetic anisotropy [17,18]. However, it remains a challenging task to realize such a weak ligand field in surface-adsorbed organometallic complexes, and the reported magnetic anisotropy energy in surface-adsorbed organometallic complexes is in the order of several meV [19-22], with the largest magnetic anisotropy energy of about 10 meV achieved on a lead substrate, where the magnetic moment is stabilized by the superconducting gap [23].

In this Letter, by a comprehensive study of low-temperature scanning tunneling microscopy/spectroscopy (STM/STS) in magnetic field, we report on magnetic anisotropy energy of more than 15 meV in an Fe-porphyrin complex on Au(111) surface. Density functional theory (DFT) calculations further identify this Fe-porphyrin complex as an intermediate of the on-surface metalation reaction, in which the Fe atom is surrounded by a weak ligand field with elongated Fe-N bonds. This weak ligand field keeps Fe in its high-spin state and preserves its large orbital angular momentum, which gives rise to easy-axis magnetic anisotropy by spin-orbit coupling.

All the experiments were conducted in an ultra-high vacuum low-temperature scanning tunneling microscope (Unisoku) with a base pressure better than $1.0\times10^{-10}$ Torr at 4.9 K. The Au(111) substrate was cleaned by cycles of $Ar^+$ sputtering and annealing. Excess Fe and submonolayer tetra-pyridyl-porphyrin [TPyP, Fig. 1(a)] molecules were deposited on Au(111) substrate held at room temperature. The d$I$/d$V$ spectra were acquired using a lock-in amplifier with a sinusoidal modulation of 987.5 Hz at 0.1 mV.

The DFT calculations were performed using projector-augmented wave (PAW)

pseudopotential and plane-wave basis set with energy cutoff at 400 eV. The van der Waals (vdW) density functional computations in conjunction with the Perdew-Burke-Ernzerhof (PBE) functional were performed to give accurate adsorption configurations. The calculation model was constructed by a Au(111)-(8×8) surface containing three Au atomic layers, in which the topmost layer was fully relaxed with the bottom two layers fixed. A vacuum region larger than 15 Å was applied. Structural optimizations adopted gamma-point-only K sampling, and all the structures were optimized until the force on each atom was less than 0.04 eV/Å. An effective Hubbard term $U$=4.0 eV was added to describe the $d$ orbitals of the Fe atoms. All the calculations were performed with Vienna Ab initio Simulation Package (VASP).

After Fe and TPyP molecules are deposited sequentially on Au(111) substrate, chain-like structures are formed on the surface [24]. We have identified three types of molecules. Figure 1(b) shows the STM image of a molecular chain containing all of them, denoted as type-I, type-II and type-III from bottom to top. All these molecules show intra-molecular patterns with $C_{2v}$ symmetry – type-I shows a depression in the center, type-II shows a depressed center and four bright lobes separated by two bisecting planes along the molecular high-symmetry axes, and type-III shows a rod-like protrusion sandwiched between two shoulder features along one high-symmetry axis. Since type-I molecules are similar to the TPyP molecules in the absence of Fe (see Supplemental Material [25]), they are attributed to free-base TPyP molecules. Type-III molecules show close resemblance to previously reported Fe(II)-porphyrin molecules synthesized through on-surface metalation reactions [26-28], *i.e.*, the final product (*f*-FeTPyP) of the metalation reaction. Type-II molecules are observed exclusively in the presence of Fe, and are identified as the intermediate of the metalation reaction (*i*-FeTPyP) as discussed in the following.

The d$I$/d$V$ spectra measured near the Fermi level at the centers of the three types of molecules are compared in Fig. 1(c), together with a spectrum measured on bare Au(111) surface for reference. The d$I$/d$V$ spectra measured on type-I (TPyP) molecule and Au(111) are featureless in the range of -80 meV to +80 meV. In contrast, the type-II molecule shows two symmetric conductance steps at ±15 meV, while type-III (*f*-FeTPyP) molecule shows multiple conductance steps.

These d*I*/d*V* spectra of type-II and type-III molecules are interpreted as inelastic electron tunneling spectroscopy (IETS), and the differential conductance steps are attributed to opening of additional inelastic tunneling channels when the energy of tunneling electrons exceeds certain thresholds. The IETS measured on type-II molecule indicates an excitation process with excitation energy of 15 meV. To unveil the origin of this excitation, we acquired IETS at the center of type-II molecules in magnetic fields perpendicular to the substrate (Fig. 2). For each IETS, the excitation energy is derived by fitting the spectrum with two temperature-broadened step functions distributed symmetrically with respect to the Fermi level:

$$\sigma(eV) = \sigma_0 + h^- \theta(\varepsilon + eV) + h^+ \theta(\varepsilon - eV), \qquad (1)$$

where $\sigma(eV)$ is the measured differential conductance at bias $V$, $e$ is the electron charge, $\sigma_0$ describes the non-zero background, $h^-$ ($h^+$) is the height of inelastic step at negative (positive) bias polarity, $\varepsilon$ is the excitation energy and

$$\theta(\varepsilon) = \frac{1+(x-1)\exp(x)}{(\exp(x)-1)^2}, \qquad (2)$$

with $x=\varepsilon/k_B T$, $k_B$ is the Boltzmann's constant and $T$ is the temperature [29].

Figure 2(a) shows two type-II molecules, denoted as A and B, in the same chain. The least-square fits of IETS measured on the two molecules are plotted in Fig. 2(b) and 2(c), respectively. Figure 2(d) shows the evolutions of the extracted excitation energies with increased magnetic field. The excitation energy varies with magnetic field, which is a signature of Zeeman splitting, and indicates the observed excitation is caused by the spin-flip process [9,11]. The excitation energy of molecule A increases from 16.7±0.1 meV at 0 T to 17.7±0.1 meV at 7 T, while the excitation energy of molecule B decreases from 18.7±0.1 meV at 0 T to 17.9±0.1 meV at 7 T. As discussed below, the inverse trends are given by different spin-flip ground states that have opposite spin directions.

To the lowest order, the spin excitation in a type-II molecule is modeled by a spin Hamiltonian [11,19,29]

$$\widehat{H} = g\mu_B \vec{B} \cdot \hat{\vec{S}} + D\hat{S}_z^2 + E(\hat{S}_x^2 - \hat{S}_y^2). \qquad (3)$$

The first term represents the Zeeman splitting caused by external magnetic field, in which $g$ is the Lande factor, $\mu_B$ the Bohr magneton, and $\hat{\vec{S}} = (\hat{S}_x, \hat{S}_y, \hat{S}_z)$ the spin operator. The second (third) term describes the axial (transverse) magnetic anisotropy, with $D$ ($E$) the axial (transverse) magnetic anisotropy parameter.

Our DFT calculations indicate the Fe atom in type-II molecule has a spin state of $S=2$ (discussed below), which is also in agreement with previous results in similar Fe-porphyrin system [24]. Diagonalization of Eq. 3 with $S_z=|+2\rangle$, $|+1\rangle$, $|0\rangle$, $|-1\rangle$ and $|-2\rangle$ yields the eigenstates and eigenenergies. By fitting the measured excitation energies to spin-flip excitations, we found that the primary anisotropy axis, *i.e.* the $z$ axis in Eq. 3, is assigned to the direction perpendicular to the molecular plane (parallel to the magnetic field) [30]. The best fit gives $D$=-5.57±0.01 meV, $E$=0.00±0.01 meV and $g$=1.98±0.01 for molecule A, and $D$=-6.23±0.01 meV, $E$=0.00±0.01 meV and $g$=1.92±0.01 for B.

The axial magnetic anisotropy parameter $D$<0 means easy-axis magnetic anisotropy perpendicular to the molecular plane. The transverse anisotropy parameter, $E$, is close to zero for both A and B, which minimizes the intermixing of different spin states, and makes $\widehat{H}$ commute with $\hat{S}_z$ when $B$ is along the $z$ direction. In this situation, $\widehat{H}$ and $\hat{S}_z$ share the same eigenstates (see Fig. 2e). Based on the equation given by C. F. Hirjibehedin *et al* [11], the relative IETS step heights for transitions in different magnetic field should be identical, which can be seen in the normalized IETS (see Supplemental Material [25]).

The fitted excitation energies are plotted in Fig. 2(d), and are in good agreement with experimental results. The spin-flip excitations in A and B are depicted by red and blue arrows in Fig. 2(e), respectively, which correspond to excitation from $|-2\rangle$ to $|-1\rangle$ with $\Delta S_z$=+1, and $|+2\rangle$ to $|+1\rangle$ with $\Delta S_z$=-1. Thus A and B are at different spin-flip ground state with opposite spin directions along $z$ axis.

At zero field, the measured zero-field splitting (ZFS) energy gives the energy difference between the ground state and an excited state, which has less spin projection on *z* axis. The measured ZFS energy also gives the magnetic anisotropy energy of a magnetic atom/molecule [3,9,11,19]. Thus molecule A has the magnetic anisotropy energy of 16.7 meV, and molecule B has the magnetic anisotropy energy of 18.7 meV. It is worth noting that the measured ZFS energies are irrespective of junction impedance, but do vary from molecule to molecule. Statistically the measured ZFS energy falls in the range of 9 to 37 meV, with the centroid from 14 to 17 meV (see Supplemental Material [25]). These values are significantly larger than reported values for single atoms or molecules on metal substrates, even exceed the magnetic anisotropy energy of single Fe atom on MgO (determined to be 14.0 meV by STM) [31].

Considering the transverse anisotropy parameter *E*=0 in the spin Hamiltonian, at zero magnetic field $|+2\rangle$ and $|-2\rangle$ states are degenerate, and separated by an energy barrier of $4|D|$ (>20 meV). The external magnetic field lifts the degeneracy between these two states, and makes $|-2\rangle$ state energetically favourable. In this case, the $|+2\rangle$ state can transfer to $|-2\rangle$ either by multiple spin-flip excitations [inset of Fig. 2(e)], or magnetic tunnelling [32-34]. Obviously, the thermal energy given by the environmental temperature of 4.9 K is insufficient to induce such multiple spin-flip excitations which have a threshold energy of 17.9 meV at 7 T (the spin excitation energy from $|+2\rangle$ to $|+1\rangle$ state). In our experiment, both the $|-2\rangle$ state in molecule A and the $|+2\rangle$ state in molecule B were stable in magnetic field of 7 T for more than 15 hours, indicating the magnetic tunnelling plays a minor role, and type-II molecules have long spin lifetimes, which may have potential applications in non-volatile data storage.

The large magnetic anisotropy of type-II molecule originates from the ligand field given by its special configuration. DFT calculations reveal that in type-II molecule the porphyrin macrocycle is non-dehydrogenated, and the Fe atom is not fully incorporated into the macrocycle. Thus the type-II molecule is an intermediate of the metalation reaction (*i*-FeTPyP), in contrast to final product (*f*-FeTPyP), in which Fe atom is fully incorporated into the dehydrogenated porphyrin macrocycle. Figures 3(a) and 3(b) compare the optimized adsorption configurations of *i*-FeTPyP and *f*-FeTPyP. The simulated STM images reproduce well the topographical features observed experimentally [Fig. 3(c)], validating the proposed structure models. As illustrated in the upper panel of Fig. 3(b), the central Fe atom in *i*-FeTPyP does not sit in the same plane as the molecular backbone, but lies between the Au substrate and the molecule plane. The central Fe atom is 1.01 Å and 1.09 Å lower than the iminic (-N=) and pyrrolic (-NH-) nitrogen atoms, respectively. In *f*-FeTPyP, Fe is further lifted up from the Au substrate by 0.66 Å, being much closer to the molecule plane [lower panel of Fig. 3(b)], which accounts for the central protrusion in STM image. The Fe-N bonds in *i*-FeTPyP, especially those with H attached (Fe-pyrrolic N ~2.43 Å, Fe-iminic N ~2.17 Å), are greatly elongated by more than 16% with respect to the Fe-N bonds in *f*-FeTPyP (~2.09 Å). As the result, the ligand field surround the Fe atom in *i*-FeTPyP is weaker than that in *f*-FeTPyP.

The calculated projected density of states (PDOS) of the *d*-orbitals of the central Fe atom in *i*-FeTPyP [upper panel of Fig. 3(d)] indicates only the $d_{xy}$ orbital, partially overlapped with $d_{x^2-y^2}$ orbital, is doubly occupied, and other four orbitals are singly occupied by electrons with the same spin direction [upper inset of Fig. 3(d)]. Due to the relatively weak in-plane ligand field, the spin-orbit coupling together with the mixture of in-plane orbitals ($d_{xy}$ and $d_{x^2-y^2}$) acts easily to restore an unquenched orbital angular momentum perpendicular to the molecule plane. Thus the Fe atom in *i*-FeTPyP is in its $^5D$ spin configuration (*L*=2, *S*=2), which can give rise to a large magnetic anisotropy energy by spin-orbit coupling [3]. Moreover, the dominant in-plane orbital motions can direct the electron spins parallel to the orbital angular momentum by spin-orbit coupling, and results in an easy-axis perpendicular to the molecular plane [19]. In contrast, the occupation of *d* orbital of the Fe atom in *f*-FeTPyP is drastically changed by its relatively strong in-plane bonds with N atoms. The doubly occupied orbital is not the mixture of $d_{xy}$ and $d_{x^2-y^2}$ but $d_{z^2}$ instead,

*i.e.*, the Fe atom is in $^5S$ term ($L=0$). Therefore, the out-of-plane orbital motions are dominant in *f*-FeTPyP [lower panel of Fig. 3(d)].

In our experiments, the maximum proportion of *i*-FeTPyP molecule was achieved to ~45% by moderate annealing at 350 K for 30 min. After annealing to 500 K, some *i*-FeTPyP molecules even survived on the surface, indicating the stability of *i*-FeTPyP molecules. It is worth to note that this reaction intermediate has not been observed before [35]. In our comparative experiment between tetra-phenyl-porphyrin (TPP) molecules and Fe atoms on Au(111) substrate, all the TPP molecules were metalated into final product *f*-FeTPP at room temperature (see Supplemental Material [25]). We also note that on Ag(111) surface TPyP molecules were metalated into *f*-FeTPyP at 320 K [28]. More work is needed to reveal the mechanism that makes *i*-FeTPyP stable at room temperature and affects the metalation reaction between TPyP molecules and Fe atoms on Au(111).

In conclusion, we have synthesized Fe-porphyrin complexes in a controlled way, which are intermediates of the on-surface metalation reaction between TPyP and Fe atoms, and achieved large magnetic anisotropy of more than 15 meV in these complexes. The IETS in magnetic field discriminated two magnetic states of opposite spin directions with long spin lifetime. Since the Fe atom is protected by molecular ligands, the sample can be processed at room or even higher temperatures. The reported system may have potential applications in non-volatile data storage.

We thank Prof. Yifeng Yang, Prof. Xingqiang Shi, Prof. Ziliang Shi, Dr. Tao Lin and Mr. Kai Li for useful discussions. This work is supported by the National Key R&D Program of China (Grants No. 2017YFA0303600 and 2016YFA0300600) and the Hundred Talents Program of the Chinese Academy of Sciences. S.M. acknowledges MOST (Grants No. 2016YFA0300902 and 2015CB921001).

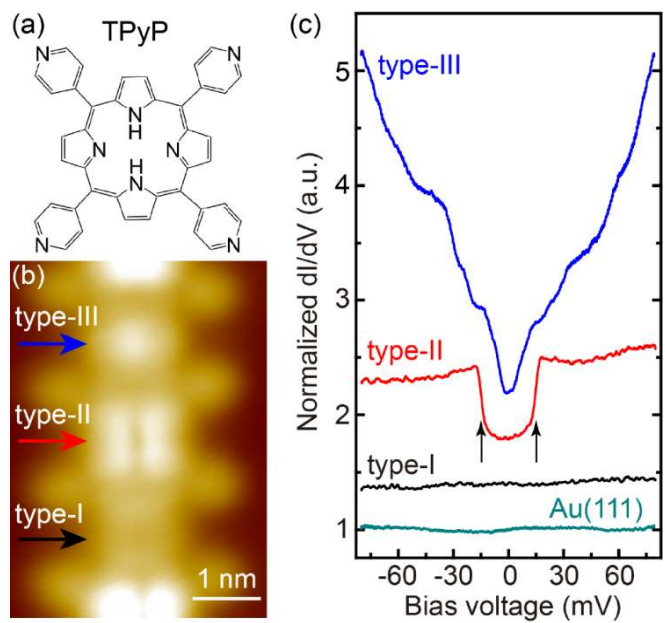

FIG. 1 (color online). (a) Chemical structure of the TPyP molecule. (b) STM image of the three types of molecules (-1.5 V, 100 pA), in which type-I, type-II and type-III molecules are indicated by black, red and blue arrows, respectively. (c) Representative d$I$/d$V$ spectra taken at the centers of the three types of molecules and bare Au(111) surface. The conductance steps in the d$I$/d$V$ spectrum of type-II molecule are indicated by arrows.

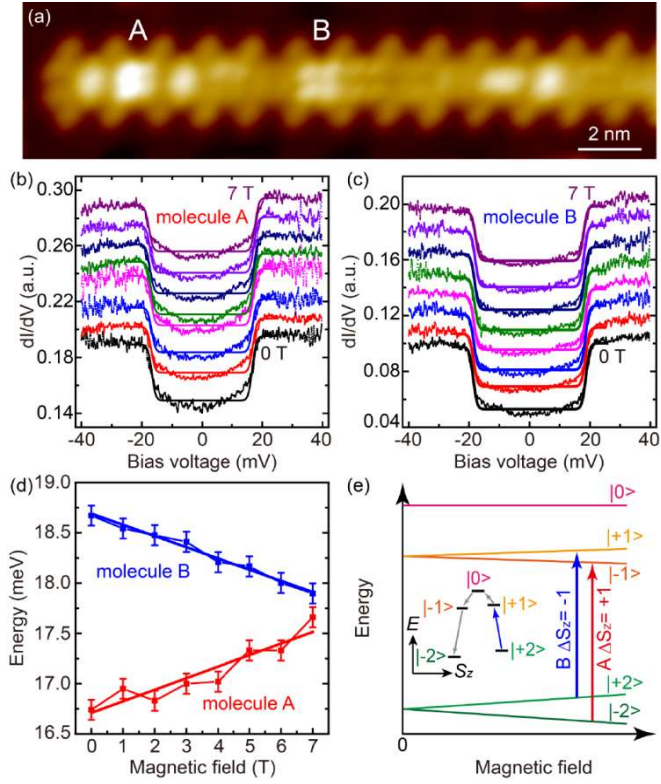

FIG. 2 (color online). (a) STM image of a molecular chain containing two type-II molecules, denoted as A and B (-1.5 V, 50 pA). (b) and (c) d$I$/d$V$ spectra taken at the centers of molecule A and B with the same tip in perpendicular magnetic field from 0 to 7 T at an interval of 1 T. The spectra are vertically shifted for clarity. The tip was stabilized at $V$=40 mV and $I$=500 pA when measuring the spectra. The fitting lines of respective spectra are plotted in solid lines of the same color. (d) Excitation energies extracted from the fitting lines in (b) and (c). The error bars are given by the amplitude of modulation during measurements. The red and blue lines are the fittings of the spin-flip excitation energy in molecule A and B, respectively. (e) Schematic of the energy-level spacing of $S$=2 states in magnetic field. The red and blue arrows indicate the spin-flip excitations in molecule A and B, respectively. Note that the exact ZFS energies of the two molecules are different, and we just show the spin-flip excitation schematically. Inset: The multiple spin-flip transitions from $|+2\rangle$ to $|-2\rangle$ state, with threshold energy defined by the blue arrow.

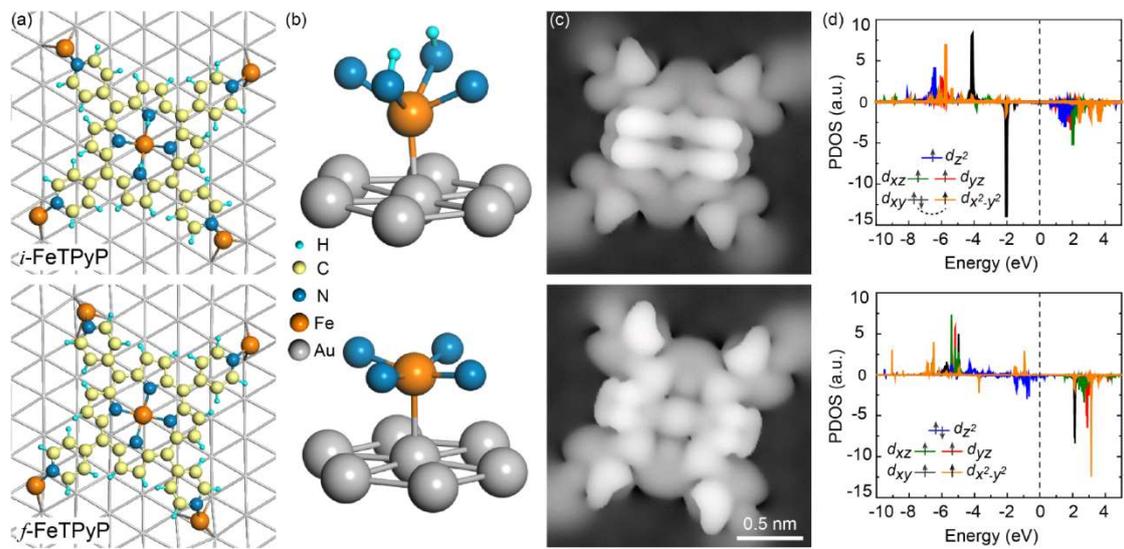

FIG. 3 (color online). (a) Optimized adsorption configurations of (upper) *i*-FeTPyP and (lower) *f*-FeTPyP on Au(111), respectively. (b) The atoms surrounding Fe in (upper) *i*-FeTPyP and (lower) *f*-FeTPyP according to (a). (c) DFT simulated STM image of (upper) *i*-FeTPyP and (lower) *f*-FeTPyP at -1.0 V. (d) Calculated spin-polarized PDOS of the *d*-orbitals of the central Fe atom in (upper) *i*-FeTPyP and (lower) *f*-FeTPyP. The occupations of the *d*-orbits are shown in the insets.